\documentclass[prd,superscriptaddress,amsfonts,amssymb,amsmath,showpacs,onecolumn,nofootinbib]{revtex4-2}
\usepackage{bm}
\usepackage{amsfonts}
\usepackage{latexsym}
\usepackage{graphicx}
\usepackage{amsmath}
\usepackage{palatino}
\usepackage{xcolor} 
\usepackage{mathpazo}
\usepackage{xcolor}
\usepackage{rotating}
\usepackage{adjustbox}
\usepackage{tensor}
\usepackage{textcomp}
\usepackage{float}
\usepackage{booktabs}
\usepackage{dcolumn}
\usepackage[titletoc]{appendix}
\usepackage{booktabs}
\usepackage{multirow}
\usepackage{hyperref}
\hypersetup{colorlinks,citecolor=blue}
\usepackage{amsmath}
\usepackage{xcolor}
\usepackage{orcidlink}
\usepackage{epsfig}
\usepackage{caption}
\usepackage{subcaption}
\usepackage{commath}
\hypersetup{colorlinks,citecolor=blue}
\hypersetup{colorlinks=true,linkcolor=magenta,filecolor=magenta,    urlcolor=blue}

\def\be{\begin{equation}}
\def\ee{\end{equation}}
\def\bea{\begin{eqnarray}}
\def\eea{\end{eqnarray}}

\begin{document}

\title{Reconstructing $f(R)$ gravity from Viaggiu Holographic Dark Energy under Hubble, Event Horizon and Granda–Oliveros Cutoffs}

\author{Arushi Jhunjhunwala}
\email{arushi.agrawal812@gmail.com}

\author{Sayani Maity}
\thanks{Corresponding author}
\email{sayani.office88@gmail.com}

\affiliation{Department of Mathematics, Sister Nivedita University,\\
DG-1/2, Action Area 1, New Town, Kolkata-700156, India.}

\begin{abstract}
This work explores the reconstruction of $f(R)$ gravity within the framework of newly proposed Viaggiu holographic dark energy (VHDE), which incorporates entropy corrections due to the dynamical nature of cosmological horizons. By implementing a correspondence between VHDE and curvature-induced energy density, we generate explicit forms of $f(R)$ for different infrared cutoffs, namely the Hubble horizon, future event horizon, and Granda–Oliveros cutoff. The resulting models are analyzed graphically through the cosmological parameters such as the equation of state and deceleration parameter, revealing a successful description of the transition from decelerated to accelerated expansion. The viability of the reconstructed models is further examined through stability conditions and local gravity constraints, that yields the nature of the models as free from ghost and tachyonic instabilities and demonstrates VHDE-inspired $f(R)$ gravity model as a consistent and flexible geometric framework for explaining the late-time cosmic acceleration of the Universe.
\end{abstract}

\maketitle

\section{Introduction}
The detection of the late-time accelerated expansion of the Universe through various observational data coming from Type Ia supernovae, measurements of the Cosmic Microwave Background (CMB) and large-scale structure surveys  
has firmly established the existence of a mysterious exotic kind of entity termed as dark energy \cite{Riess1998,Perlmutter1999,Planck2018}. This is the dominant component of the entire cosmic budget governing the present cosmic dynamics. Although the cosmological constant remains the simplest candidate to explain this acceleration, it is plagued by serious theoretical issues such as the fine-tuning and coincidence problems, thereby motivating the exploration of modifications of Einstein's field equations \cite{Weinberg1989,Sahni2000}.
The first approach preserves the framework of General Relativity (GR) by introducing an additional energy component into the right-hand side of the Einstein field equations.  These unresolved issues have stimulated the development of various dynamical dark energy scenarios based on scalar fields, exotic fluids, and models inspired by quantum gravity \cite{COPELAND_2006}. The second approach attributes the accelerated expansion not to an additional cosmic component but to a modification of the gravitational interaction itself. In this case, the matter content of the Universe remains unalterd, while the Einstein--Hilbert action or the underlying theory of gravity is generalized, yielding a variety of modified gravity theories \cite{NOJIRI_2007}. Consequently, the observed cosmic acceleration is interpreted as a manifestation of the geometric properties of spacetime rather than the effect of an unknown form of energy.\\

Among the various dynamical dark energy scenarios, Holographic Dark Energy (HDE), inspired by the holographic principle, has emerged as a promising framework in which the dark energy density is determined by an infrared (IR) cutoff scale, allowing for a dynamical description of cosmic acceleration \cite{Li2004}.
The holographic principle suggests that the number of degrees of freedom of a physical system encoded on its boundary and scales with the boundary area instead of the bulk volume.. Based on this principle, Cohen \textit{et al.} proposed that the vacuum energy density should satisfy a relationship between the ultraviolet (UV) and infrared (IR) cutoffs, leading to the standard holographic dark energy density $\rho_D \propto L^{-2}$, where $L$ denotes the infrared cutoff length. Although the standard HDE model successfully explains the late-time accelerated expansion of the Universe, several theoretical and observational issues, including the choice of a suitable infrared cutoff and the nature of the dark energy equation of state, have motivated the development of generalized holographic dark energy models.\\

One of the earliest extensions is the Tsallis holographic dark energy (THDE) model \cite{MSDS19, SARDAR2025169891,doi:10.1142/S0219887820501704}, which is based on the non-extensive Tsallis entropy. Since gravitational systems exhibit long-range interactions and non-additive thermodynamic properties, Tsallis entropy provides a natural generalization of the standard Bekenstein-Hawking entropy. In this framework, the dark energy density acquires the form $\rho_D \propto L^{2\delta-4}$, where $\delta$ is the non-extensive parameter. The additional degree of freedom introduced by $\delta$ enriches the cosmological dynamics and allows the model to accommodate both quintessence and phantom regimes. Subsequently, the Rényi holographic dark energy (RHDE) \cite{MSDS19,SARDAR2025169891,doi:10.1142/S0219887820501704} model was proposed by employing Rényi entropy, which can be viewed as the extensive counterpart of Tsallis entropy. The resulting dark energy density contains logarithmic corrections arising from generalized thermodynamics and exhibits a dynamical behavior that can closely mimic the $\Lambda$CDM model at late times. Compared with THDE, RHDE often demonstrates improved thermodynamic consistency and stability properties, making it an attractive alternative for describing cosmic acceleration. A more general framework was introduced through the Sharma-Mittal holographic dark energy (SMHDE) model \cite{MSDS19,SARDAR2025169891,doi:10.1142/S0219887820501704}. Sharma-Mittal entropy unifies both Tsallis and Rényi entropies through two independent parameters and therefore provides a broader thermodynamic foundation for holographic cosmology. Recently, significant attention has been devoted to Barrow holographic dark energy (BHDE), which originates from Barrow's proposal that quantum-gravitational effects may induce a fractal deformation of the black-hole horizon. In this scenario, the horizon entropy is modified by the Barrow exponent $\Delta$, leading to a generalized dark energy density of the form $\rho_D \propto L^{\Delta-2}$. The BHDE model provides a direct connection between holographic dark energy and quantum gravity corrections, and it has been shown to generate a rich cosmological phenomenology, including quintessence-like behavior, phantom evolution, and crossings of the phantom divide.
To further explore the implications of Barrow entropy, several generalized versions collectively known as New Barrow holographic dark energy (NBHDE) models have been proposed. These models incorporate generalized infrared cutoffs, modified entropy-area relations, or evolving Barrow exponents, thereby extending the cosmological flexibility of the original BHDE framework. Such extensions have been found to provide better agreement with various cosmological observations while retaining the essential quantum-gravity-inspired features of Barrow entropy.
Particular attention has been given to entropy corrected holographic model, where modifications to the horizon entropy can significantly alter the dark energy dynamics. In this direction, the recently proposed Viaggiu holographic dark energy (VHDE) model, based on generalized entropy formulation of Viaggiu, has attracted interest due to its ability to capture the effects of the dynamical nature of cosmological horizons and to generate cosmological behaviour that differ from conventional holographic dark energy scenarios \cite{Saha_2026}. More recently, in reference \cite{Halder_2026}, the cosmological viability of the VHDE model has been explored using late-time observational data. The constraints of the model parameters are obtained by employing most recent  Baryon Acoustic Oscillation (BAO) measurements from DESI DR2, DES-Dovekie compilations, together with Cosmic Chronometers (CC), Type Ia supernova samples from the PantheonPlus, Union3.0. \\

Among various modified gravity models proposed in literature $f(R)$ gravity model \cite{COPELAND_2006} emerged one of the most extensively studied framework providing an alternative geometric explanation for the accelerated expansion. In this framework, the Ricci scalar in the Einstein–Hilbert action is generalized to a function $f(R)$, and the resulting curvature effects can effectively mimic dark energy behavior. A powerful approach that connects dark energy models with modified gravity is the reconstruction technique, in which one starts from a given dark energy scenario and derives the corresponding form of $f(R)$ capable of reproducing the same cosmological evolution.
Reconstruction technique have been extensively employed to establish correspondences between dark energy models and modified gravity theories. For instance, $f(R)$ gravity reconstructed from Barrow holographic dark energy has been shown to exhibit cosmological features, including transition between dark energy regimes and consistency with thermodynamic requirements \cite{Sarkar:2021izd}. More recently, the reconstruction scheme of $f(R)$ gravity model from Barrow agegraphic and new barrow agegraphic dark energy has been studied in \cite{MAITY2025102184}and also the authors have generated the   parameter constraints employing the observational data. In reference \cite{doi:10.1142/S0219887824502487}, the reconstruction of $f(T)$ gravity model has been explored from parameter constraints from the Modified Chaplygin–Jacobi and Modified Chaplygin–Abel gas dark energy models. In reference \cite{doi:10.1142/S0217732322502042}, ordinary holographic dark energy, ordinary new agegraphic dark energy, entropy-corrected holographic dark energy in power-law and logarithmic versions and entropy-corrected new agegraphic dark energy in power-law and logarithmic versions with two classes of the scale factor have been considered to explore the reconstruction scheme of gravity $f(P)$ cubic gravity model. In reference \cite{Hamani_Daouda_2012},  a model of f(T) gravity has been developed according to the holographic dark energy model. The reconstruction of $f(Q,T)$ gravity from Barrow holographic dark energy with different infrared cutoffs has been done in \cite{Mohanty:2026laq}, where the reconstructed model was further constrained through the effective gravitational coupling. Likewise, in \cite{Behera:2026yya} accelerating nonlinear $f(T)$ gravity models have been reconstructed taking a hybrid scale factor, and their cosmological dynamics have been explored with Bayesian statistical analyses. More recently, Barrow holographic dark energy has also been reconstructed within the framework of $f(Q,T)$ gravity, and the resulting models have been tested against cosmological observations in \cite{Zhang:2025oki}. These studies demonstrate that the reconstruction technique provides a powerful framework for establishing viable modified gravity models consistent with both theoretical expectations and observational data.\\

Motivated by these developments, in this work we reconstruct the $f(R)$ gravity model corresponding to the VHDE scenario by considering different choices of the infrared cutoff, namely the Hubble horizon cutoff, the future event horizon cutoff, and the Granda–Oliveros cutoff. This allows us to examine how different IR cutoffs influence the reconstructed gravitational dynamics and to explore the cosmological viability of VHDE-inspired modified gravity models. In this way, the late-time acceleration of the Universe can be interpreted within a unified geometric framework, providing deeper insight into the interplay between holographic dark energy and modified gravity \cite{viaggiu2019holographic,Nojiri2006}. Our work is organized as follows: in section II, we present basic formalism of modified $f(R)$ gravity. The reconstruction scheme of the $f(R)$ model from VHDE by considering three different IR cutoffs are executed in section III. Section IV is devoted to the cosmological analysis of the reconstructed models via evolution of the equation of state parameter and the deceleration parameter. A comprehensive viability analysis is also performed by Positive Effective Gravitational Coupling, Stability Condition and Solar System Constraints tests. Finally the main findings and concluding remarks are summarized in section V.

\section{Basic Equations of Modified $f(R)$ Gravity Model}\label{sec_2}

The dynamical evolution of the Universe across different cosmological epochs is governed by the properties of the cosmic fluid responsible for its expansion. In order to describe these phases more accurately, several extensions of general relativity have been proposed. Among them, the $f(R)$ gravity framework has emerged as one of the most compelling modified gravity theories.
The action for $f(R)$ gravity is reads \cite{RevModPhys.82.451, MAITY2025102184}
\begin{equation}\label{2.1}
S=\frac{1}{2\kappa^2}\int d^4x\sqrt{-g}\left[ R+f(R)+2\kappa^2 L_m \right],
\end{equation}
where $g$, $L_m$, $G$, $R$ represents the determinant of the metric tensor $g_{\mu\nu}$, the matter Lagrangian density, the gravitational constant and the Ricci scalar respectively. By performing variation of the action (\ref{2.1}) with respect to the metric tensor $g_{\mu\nu}$, the modified field equations can be obtained as
\begin{equation}\label{2.2}
R_{\mu\nu}-\frac{1}{2}R g_{\mu\nu}-\frac{1}{2}g_{\mu\nu}f(R)+R_{\mu\nu}f^{'}(R)+\left(g_{\mu\nu}\Box-\nabla_\mu \nabla_\nu \right)f^{'}(R)=\kappa^2 T_{\mu\nu}.
\end{equation}
The form of the energy-momentum tensor for a perfect fluid distribution is $T_{\mu\nu}=\text{diag}(-\rho,p,p,p)$, where $R_{\mu\nu}$ is the Ricci tensor and $f^{'}(R)=\frac{df}{dR}$. The standard Einstein field equations are recovered in the limit $f(R) \rightarrow 0$.\\

Here we consider a spatially flat Friedmann–Lemaître–Robertson–Walker (FLRW) model of the Universe described by the metric:
\begin{equation}\label{2.3}
ds^2=-dt^2+a^2(t)\sum_{j=1}^{3} dx^i dx^j.
\end{equation}
Under this assumption, the modified field equations reduce to the Friedmann equations as follows:
\begin{equation}\label{2.4}
3H^2=\kappa^2 \rho+ \rho_{f(R)},
\end{equation}

\begin{equation}\label{2.5}
2\dot{H}+3H^2=-\kappa^2 p+ p_{f(R)},
\end{equation}
where the effective energy density and pressure arising from curvature contributions are given by
\begin{equation}\label{2.6}
\rho_{f(R)}= \frac{1}{\kappa^2}\left[3 f^{'}(R)(\dot{H}+H^2)-\frac{1}{2}f(R)-18f^{''}(R)(4H^2\dot{H}+H\ddot{H}) \right],
\end{equation}
\begin{eqnarray*}
p_{f(R)}= -\frac{1}{\kappa^2}\Big{[}\frac{1}{2}f(R)-(\dot{H}+3H^2)f^{'}(R)-6\left(8 H^2\dot{H}+6H\ddot{H}+ 4\dot{H}^2+\ddot{H}\right)f^{''}(R)
\end{eqnarray*}
\begin{equation}\label{2.7}
+36\left( H \ddot{H}+4\dot{H}^2 \right)^2 f^{'''}(R) \Big{]}.
\end{equation}

The Ricci scalar for the FLRW spacetime is expressed as
\begin{equation}\label{2.8}
R=6(\dot{H}+2H^2).
\end{equation}

The additional terms $\rho_{f(R)}$ and $p_{f(R)}$ originate from the modification $f(R)$ to the Einstein–Hilbert action and can be interpreted as effective curvature-induced energy density and pressure. These contributions can mimic dark energy behavior and are capable of driving both early-time inflation and late-time cosmic acceleration without introducing extra scalar fields. The conservation of the energy-momentum tensor yields the continuity equation, given by:
\begin{equation}
    \dot\rho + 3H(\rho + p)=0
\end{equation}
This equation governs the evolution of the energy density of the cosmic fluid in an expanding Universe. It ensures that the total energy is conserved as the Universe evolves.
The total energy density of the Universe is assumed to consist of two components, namely matter and dark energy, and can be expressed as:
 \begin{equation}
     \rho = \rho_m + \rho_d
 \end{equation}
where $\rho_m$ represents the energy density of the dark matter and $\rho_d$ denotes the effective dark energy density arising from the curvature contributions of the $f(R)$ gravity model.\\

For the non-interacting case, the equation of continuity leads to:
\begin{equation}\label{2.1}
    \dot \rho_m + 3 H \rho_m = 0
\end{equation}
and
\begin{equation}
    \dot{\rho_d} + 3 H (\rho_d + p_d)=0.
\end{equation}
Exploiting equation (\ref{2.1}), we get: 
\begin{equation}
    \rho_m = \rho_{m0}(1+z)^3,
\end{equation}
where the relation between the scale factor and redshift is given by
\begin{equation}
1+z=\frac{1}{a(t)}.
\end{equation}
Here we adopt a power-law form of the scale factor as \cite{PhysRevD.74.086009}
\begin{equation}\label{2.9}
a(t)=a_0 t^n, \quad n>0,
\end{equation}
where $a_0$ and $n$ denote the present value of the scale factor and the power-law index, respectively. This choice corresponds to a Universe originating from a Big Bang singularity while avoiding future-time singularities. 
Investigating VHDE within the framework of \(f(R)\) gravity provides a natural way to unify entropy-based dark energy models with geometrically motivated modifications of gravity, both of which offer viable explanations for the late-time accelerated expansion of the Universe. This has been widely employed in the literature as an effective cosmological ansatz \cite{MAITY2025102184, SMASPR26}. Recently, this form of scale factor has been employed in \cite{Sultana_2026} for reconstruction of Tsallis Holographic Dark Energy in the background of Modified Non-Metric Gravity $f(Q,C)$. Using this ansatz, the Hubble parameter and its derivatives take the form
\begin{equation}\label{2.10}
H=\frac{n}{t}, \quad \dot{H}=-\frac{n}{t^2}, \quad \ddot{H}=\frac{2n}{t^3},
\end{equation}
and the Ricci scalar becomes
\begin{equation}\label{2.11}
R=\frac{2n^2-n}{t^2}.
\end{equation}

To provide a convenient characterization of the cosmic expansion history, we introduce the dimensionless Hubble parameter $E(Z)=\frac{H(z)}{H_0}$. The quantity $E^2(z)$ is obtained as a function of the reconstructed $f(R)$ model and its higher-order derivatives, encoding the effects of modified gravity on the evolution of the Universe and is given as:
\begin{eqnarray*}
    E^2(z) =\Big{[}\Omega_{m0} (1+z)^3 + \frac{1}{H^2_0}(-n)(1+z)^\frac{2}{n} f'(R) - \frac{1}{6}f(R) \Big{]}
\end{eqnarray*}
\begin{equation} \label{E_sq}
    \Big{[}1 - f'(R) - 24 (1+z)^\frac{2}{n} f''(R) + 12 (1+z)^\frac{2}{n} f'''(R)  \Big{]}^{-1},
\end{equation}
 where, $\Omega_{m0}$ denotes the present value of the matter density parameter, representing the fractional contribution of matter to the total energy density of the Universe at the current epoch.

\section{Reconstruction Scheme of $f(R)$ gravity model from Viaggiu Holographic Dark Energy }

The Viaggiu holographic dark energy (VHDE) model has recently emerged as a member of the holographic framework for explaining the late-time cosmic acceleration.
In this section, we aim to reconstruct the functional form of $f(R)$ gravity corresponding to the VHDE model \cite{Saha_2026}. The foundation of holographic dark energy originates from the work of Cohen \cite{Cohen:1998zx}, who established a connection between the ultraviolet (UV) and infrared (IR) cutoffs in the framework of effective quantum field theory. This relation arises due to the constraint imposed by black hole formation. Specifically, if $\rho_d$ represents the quantum zero-point energy density associated with a short-distance (UV) cutoff, then the total energy contained within a region of size $L$ must be within the range of the mass of a black hole of the same size not beyond that. This fact leads to the inequality \cite{Li2004,Wang:2017xyh,Bekenstein:1981jp}
\begin{equation}
M_p^{-2} L^4 \rho_d \leq S,
\end{equation}
where $L$ denotes the horizon radius and $S$ is the corresponding entropy. In the context of the Viaggiu entropy, the entropy-area relation is modified by incorporating an additional volume-dependent contribution. Expressed in terms of the length scale $L$, the entropy takes the form
\begin{equation}
S = \pi L^2 + 2\pi H L^3,
\end{equation}
where the first term corresponds to the standard area contribution, while the second term arises due to the expansion of the Universe through the Hubble parameter $H$. This expression follows from the fact that the surface area and volume enclosed by a spherical horizon of radius $L$ are given by $4\pi L^2$ and $\frac{4}{3}\pi L^3$, respectively.

Assuming that the above inequality is saturated, one obtains the expression for the dark energy density as
\begin{equation}\label{5.21}
\rho_d = \frac{\delta^2}{8\pi} L^{-4} S 
= \frac{\delta^2}{8\pi} L^{-4} \left( \pi L^2 + 2\pi H L^3 \right),
\end{equation}
where $\delta^2$ is a positive parameter introduced to absorb possible numerical factors that are not explicitly accounted for. In \cite{Saha_2026} the authors have investigated the cosmological implications of VHDE model with the Hubble horizon and the future event horizon as infrared (IR) cut-offs. For the reconstruction of $f(R)$ gravity model we are considering these to IR cut-offs and also we are introducing VHDE with Granda Oliver (GO) cut-off.

\subsection{ Reconstruction of $f(R)$ for Viaggiu HDE with Hubble Horizon }
Furthermore, by identifying the IR cutoff with the Hubble horizon, i.e., $L = H^{-1}$, eq (\ref{5.21}) simplifies to \cite{Saha_2026}
\begin{equation}\label{5.22}
\rho_d = \frac{3\delta^2}{8\pi} H^2.
\end{equation}

Exploiting the equation(\ref{2.10}) and (\ref{2.11}) the energy density of Viaggiu with Hubble cutoff, we get:
\begin{equation}\label{HH_rho}
    \rho_d=\frac{3\delta^2nR}{8(2n-1)}
\end{equation}

Now for the reconstruction scheme, we consider the correspondence of the energy density of $f(R)$ gravity model and the energy density of VHDE with Hubble horizon cut-off (\ref{HH_rho}) that results in the following differential equation:
\begin{equation}\label{HH_f}
    96R^2f''(R)+8(n-1)Rf'(R)-\frac{4}{3}(2n-1)f(R)=n\delta^2R.
\end{equation}
Solving the differential equation (\ref{HH_f}), we get
\begin{equation}\label{5.25}
    f(R)=A_1R^{m_1}+A_2R^{m_2}+\frac{3n\delta^2R}{4(4n-5)}
\end{equation}
where
\begin{equation}\label{m1}
    m_1=\frac{13-n}{24}+\frac{1}{6\sqrt{2}}\sqrt{\frac{(n-13)^2}{8}+2n-1}
\end{equation}
and
\begin{equation}\label{m2}
    m_2=\frac{13-n}{24}-\frac{1}{6\sqrt{2}}\sqrt{\frac{(n-13)^2}{8}+2n-1}.
\end{equation}

Exploiting equations (\ref{HH_f}) and (\ref{E_sq}), we get the expression $E^2(z)$ as:
\begin{eqnarray*}
   E^2(z)=\frac{1}{ 6 H_0^2}\Big{[}6 H_0^2 \Omega_{m_0} (z+1)^3-A_1 n^{m_1} \left(6 m_1+2 n-1\right) (2 n-1)^{m_1-1} (z+1)^{\frac{2 m_1}{n}}
\end{eqnarray*}
\begin{eqnarray*}
   -A_2 n^{m_2} \left(6 m_2+2 n-1\right)  (2 n-1)^{m_2-1} (z+1)^{\frac{2 m_2}{n}} -\frac{3 \delta ^2 n^2 (2 n+5) (z+1)^{2/n}}{4 (4 n-5)}\Big{]}
\end{eqnarray*}
\begin{eqnarray*}
  \Big {[}1 -A_1 m_1 n^{m_1 -2} \left(12 m_1+n-12\right)  (2 n-1)^{m_1-1} (z+1)^{\frac{2(m_1-1)}{n}}
\end{eqnarray*}
\begin{equation}
    -A_2 m_2 n^{m_2 -2} \left(12 m_2+n-12\right)
    (2 n-1)^{m_2-1} (z+1)^{\frac{2(m_2-1)}{n}}+\frac{3 \delta ^2 n}{20-16 n}\Big{]}^{-1}
\end{equation}

\begin{figure}[htbp]
    \centering
    \begin{subfigure}[b]{0.32\textwidth}
\includegraphics[width=\linewidth]{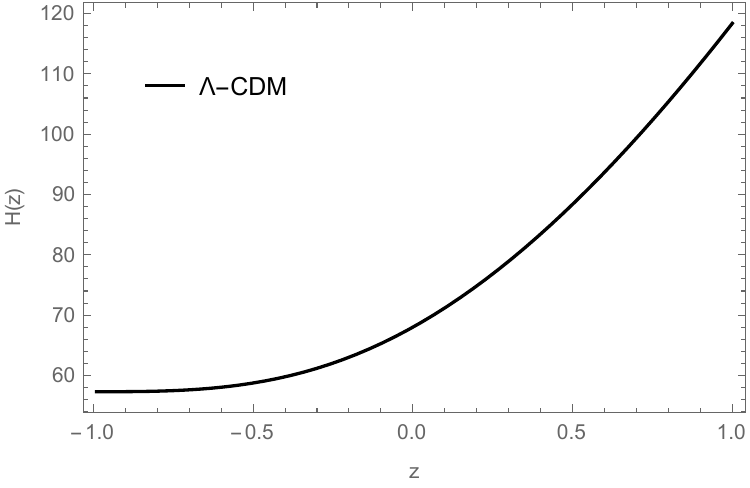}
        \caption{}
    \end{subfigure}
    \hfill
    \begin{subfigure}[b]{0.32\textwidth}
\includegraphics[width=\linewidth]{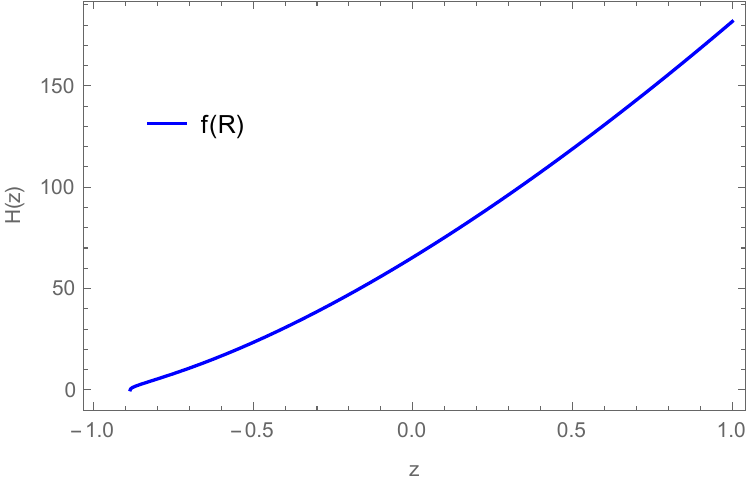}
        \caption{}
    \end{subfigure}
    \hfill
    \begin{subfigure}[b]{0.32\textwidth}
 \includegraphics[width=\linewidth]{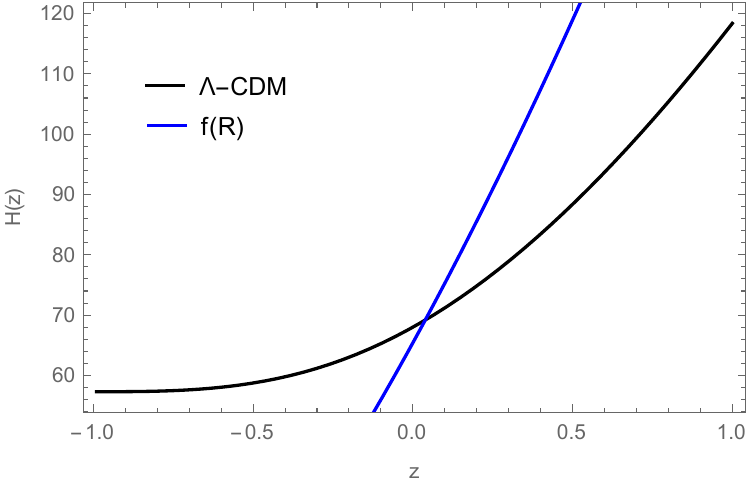}
        \caption{}
    \end{subfigure}
    
    \caption{$E^2(z)$ vs $z$ for (a) $\Lambda CDM$ model (b) Hubble Horizon, and (c) Comparision between the two }
    \label{H_vs_z}
    
\end{figure}

 Fig:(\ref{H_vs_z}), illustrates that the reconstructed $f(R)$ model closely matches the $\Lambda CDM$ behavior at present, indicating that it provides a realistic and feasible description of the Universe’s expansion.\\


\subsection{ Reconstruction of $f(R)$ for Viaggiu HDE with Future event horizon cut-off }
Furthermore, by identifying the IR cutoff with the future event Hubble horizon \cite{Faraoni:2015ula}, 
\begin{equation}\label{5.30}
  L =R_E=a(t)\int_t^\infty \frac{dt}{a(t)}=a\int_a^\infty \frac{da}{Ha^2}= a\int_x^\infty \frac{dx}{Ha},  
\end{equation}
 
and exploiting (\ref{5.30}) the expression (\ref{5.21}) simplifies to
\begin{equation}\label{5.32}
\rho_d = \frac{\delta^2}{8R_E^2}(1+2HR_E).
\end{equation}
and consequently we get the energy density of Viaggiu with future horizon as IR cut-off as follows:
\begin{equation}\label{FE_roh}
    \rho_d = \frac{\delta^2R^2(n-1)^3(3n-1)}{8n^2(2n-1)^2}
\end{equation}
Now, to implement the reconstruction scheme, we equate the energy density of the $f(R)$ gravity model and that of the VHDE model with the future event horizon as the cut-off, which yields the following differential equation:

\begin{equation} \label{5.33}
    288R^2f''(R)+24(n-1)Rf'(R)-4(2n-1)f(R)=\frac{\delta(n-1)^3(3n-1)}{n^2(2n-1)}R^2
\end{equation}

Solving the above differential equation (\ref{5.33}), we get
\begin{equation} \label{FE_f}
    f(R)=A_1R^{m_1}+A_2R^{m_2}+\frac{(n-1)^3(3n-1)\delta R^2}{4n^2(2n-1)(151-8n)}
\end{equation}
where $m_1$ and $m_2$ are given by (\ref{m1}) and (\ref{m2}) respectively. Using equations (\ref{FE_f}) and (\ref{E_sq}), we get the expression $E^2(z)$ as:
\begin{eqnarray*}
   E^2(z)=\frac{1}{ 6 H_0^2}\Big{[}6 H_0^2 \Omega_{m_0} (z+1)^3-A_1 n^{m_1} \left(6 m_1 +2n -1\right) (2 n-1)^{m_1-1} (z+1)^{\frac{2 m_1}{n}}
\end{eqnarray*}

\begin{eqnarray*}
   -A_2 n^{m_2} \left(6 m_2+2 n-1\right)  (2 n-1)^{m_2-1} (z+1)^{\frac{2 m_2}{n}} -\frac{ \delta (n - 1)^3 (3 n-1)(11+2n) (z+1)^{2/n}}{4 (151 -8 n)}\Big]
\end{eqnarray*}

\begin{eqnarray*}
  \Big {[}1 -A_1 m_1 n^{m_1 -2} \left(12 m_1+n-12\right)  (2 n-1)^{m_1-1} (z+1)^{\frac{2(m_1-1)}{n}}
\end{eqnarray*}
\begin{equation}
    -A_2 m_2 n^{m_2 -2} \left(12 m_2+n-12\right)
    (2 n-1)^{m_2-1} (z+1)^{\frac{2(m_2-1)}{n}}-\frac{(n-1)^4 (3n-1) \delta n}{2 n^2 (151-8n)}(z+1)^\frac{2}{n}\Big{]}^{-1}
\end{equation}

\subsection{ Reconstruction of $f(R)$ for Viaggiu HDE with Granda Oliver (GO) cut-off}

Here we consider the Granda--Oliveros (GO) cutoff \cite{Granda:2008dk,Granda:2009fh}, which is constructed purely from local cosmological parameters. This IR cutoff was originally proposed in Ref.~\cite{Granda:2008dk} as an infrared cutoff for the holographic dark energy density, and was further developed and generalized in Ref.~\cite{Granda:2009fh}, where additional terms involving the Hubble parameter and its time derivative were introduced to address the difficulties faced by the Hubble-horizon cutoff in producing accelerated expansion. In the GO approach, the infrared (IR) cutoff is defined as a function of the Hubble parameter and its derivative as \cite{Granda:2008dk,Granda:2009fh,MOTAGHI2024101710}
\begin{equation}\label{3.3.1}
  L_{GO} = (\alpha H^2 + \beta \dot H)^{-\frac{1}{2}},  
\end{equation}
where $\alpha$ and $\beta$ are dimensionless constants. This cutoff has since been widely employed in extended holographic dark energy frameworks based on generalized entropies, such as Tsallis and Barrow holographic dark energy \cite{Oliveros:2022qxi}.
Using (\ref{3.3.1}), the Viaggiu holographic dark energy density takes the form:
\begin{equation} \label{3.3.2}
    \rho_d = \frac{\delta^2 R \sqrt{\alpha n - \beta}}{(2n-1)}(\sqrt{\alpha n - \beta} + 2 \sqrt{n})
\end{equation}
This choice enables a causality-consistent formulation and provides a suitable setting for reconstructing f(R) gravity corresponding to Viaggiu holographic dark energy. The resulting framework naturally incorporates entropy corrections and offers a unified description of dark energy and modified gravity, thereby yielding a flexible and potentially viable model for explaining the late-time acceleration of the Universe. Now, to implement the reconstruction scheme, we equate the energy density of the $f(R)$ gravity model and that of the VHDE model with the GO as the cut-off (\ref{3.3.2}), which yields the following differential equation:

\begin{equation} \label{3.3.3}
    288R^2f''(R)+24(n-1)Rf'(R)-4(2n-1)f(R)= \frac{\delta^2 \sqrt{\alpha n - \beta }(\sqrt{\alpha n - \beta} + 2 \sqrt n)}{288}R.
\end{equation}

Solving the eq(\ref{3.3.3}), we get
\begin{equation}
    f(R)=A_1R^{m_1}+A_2R^{m_2}+ \frac{\delta^2 \sqrt{\alpha n - \beta}(\sqrt{\alpha n - \beta} + 2\sqrt{n})}{4(65 + 4 n)}R,
\end{equation}
where $m_1$ and $m_2$ are given by (\ref{m1}) and (\ref{m2}) respectively. In this case we have obtained the expression of $E^2(z)$ as

\begin{eqnarray*}
   E^2(z)=\frac{1}{6 H_0^2}\Big{[}6 H_0^2 \Omega_{m_0} (z+1)^3-A_1 n^{m_1} \left(6 m_1 +2n -1\right) (2 n-1)^{m_1-1} (z+1)^{\frac{2 m_1}{n}}
\end{eqnarray*}

\begin{eqnarray*}
   -A_2 n^{m_2} \left(6 m_2+2 n-1\right)  (2 n-1)^{m_2-1} (z+1)^{\frac{2 m_2}{n}} -\frac{\delta^2 \sqrt{\alpha n - \beta}(10 n^{\frac{3}{2}} + 4 n^{\frac{5}{2}} + (2n+5)\sqrt{\alpha n - \beta})}{4(4 n + 65)}\Big{]}
\end{eqnarray*}

\begin{eqnarray*}
  \Big {[}1 -A_1 m_1 n{^{m_1 -2}}\left(12 m_1+n-12\right)  (2 n-1)^{m_1-1} (z+1)^{\frac{2(m_1-1)}{n}}-A_2 m_2 n^{m_2 -2} \left(12 m_2+n-12\right)
\end{eqnarray*}

\begin{equation} 
    (2 n-1)^{m_2-1} (z+1)^{\frac{2(m_2-1)}{n}} -\frac{\delta ^2 \left(\alpha n-\beta +2 \sqrt{n} \sqrt{\alpha n-\beta }\right)}{4 (4 n+65)}\Big{]}^{-1}
\end{equation}


\section{Cosmological analysis}

To understand the dynamical behaviour of the reconstructed f(R) gravity model, we analyse graphically the evolution of equation of state parameter$\omega$ and deceleration parameter $q(z)$ as a function of redshift $z$ .Due to the complexity of the analytical expressions, the expressions of the EoS and the deceleration parameter are not mentioned here.

\subsection{Equation of state:}
\begin{figure}[htbp]
    \centering
    \begin{subfigure}[b]{0.32\textwidth}
\includegraphics[width=\linewidth]{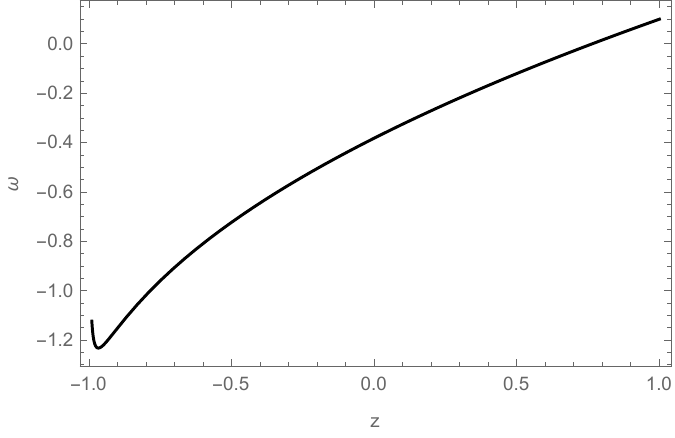}
        \caption{}
    \end{subfigure}
    \hfill
    \begin{subfigure}[b]{0.32\textwidth}
 \includegraphics[width=\linewidth]{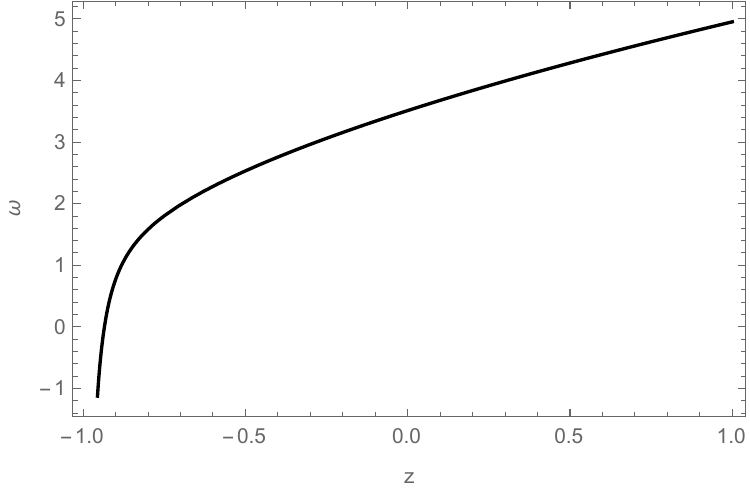}
        \caption{}
    \end{subfigure}
    \hfill
    \begin{subfigure}[b]{0.32\textwidth}
 \includegraphics[width=\linewidth]{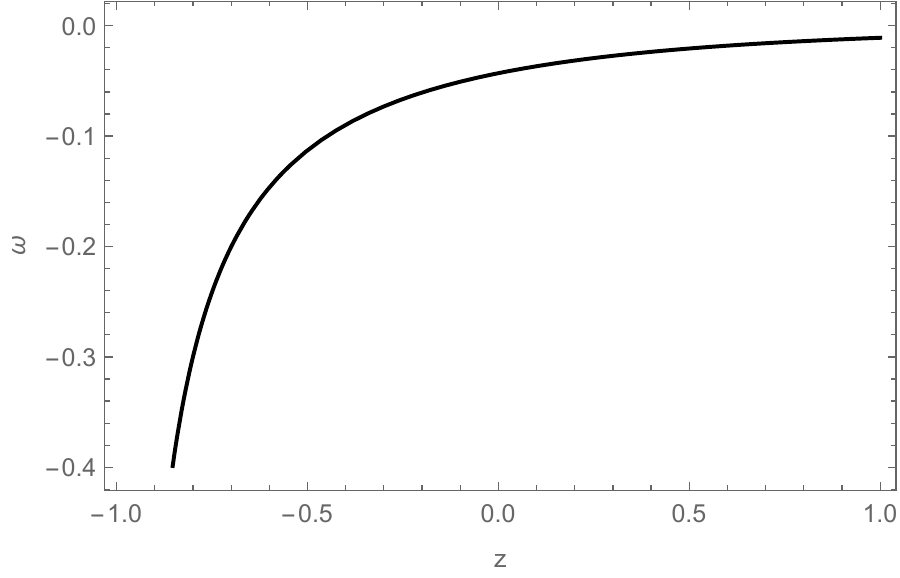}
        \caption{}
    \end{subfigure}
    
    \caption{Equqtion of state vs redshift for three reconstructed $f(R)$ models for (a)Hubble Horizon, (b)Future Event Cutoff and (c)Granda-Oliveros }
    \label{w_vs_z}
    
\end{figure}

The equation of state (EoS) parameter, defined as
\begin{equation}
\omega_d=\frac{p_d}{\rho_d}.
\end{equation}
Its value provides insight into the evolution of cosmic acceleration by identifying different cosmological phases. Specifically, $\omega_d=0$ corresponds to a dust dominated era, $-1<\omega_d<-1/3$ represents the quintessence regime, $\omega_d=-1$ describes the cosmological constant ($\Lambda$CDM) model, while $\omega_d<-1$ characterizes the phantom regime. A transition of the EoS parameter across the phantom divide line, $\omega_d=-1$, from $\omega_d>-1$ to $\omega_d<-1$ is termed as the quintom behavior, indicating a dynamical evolution of dark energy. The behavior of the equation of state parameter $\omega(z)$ is presented in Fig:(\ref{w_vs_z}) for the reconstructed models.
It is observed from Fig: \ref{w_vs_z} (a) that for the Hubble Horizon cutoff the trajectory of the EoS parameter exhibits a transition from  quintessence region to phantom region by crossing the phantom divide line which reflects a quintom-like behavior. The parameter values used are $A_1 = 0.23$, $A_2 = 0.34$, $\delta = 1.222$, and $n = 3.56$. 
In case of Future Event Horizon parameter values used are $A_1 = 21$, $A_2 = 0.002$, $\delta = 1.56$, and $n = 3.5688$. Fig~\ref{w_vs_z}(b) reflects a transition from the dust dominant era to quintessence region crossing the phantom divide line, which depicts quintom like behaviour leading to Big Rip for the Future Event Horizon. While in case of the Granda--Oliveros Cutoff, using $A_1 = 0.0005$, $A_2 = -12$, $\delta = 3.56$, $n = 3.56$, along with $\alpha = 10$ and $\beta = 5$, Fig:\ref{w_vs_z} (c) shows only the quintessence kind of behaviour throughout the evolution.

\subsection{Decceleration parameter:}
\begin{figure}[htbp]
    \centering
    \begin{subfigure}[b]{0.32\textwidth}
\includegraphics[width=\linewidth]{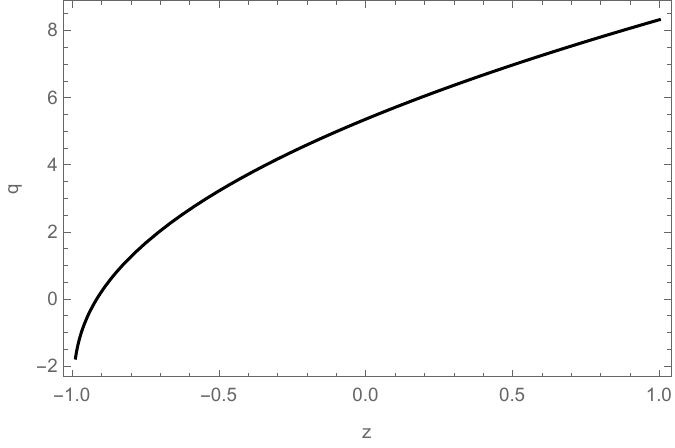}
        \caption{}
    \end{subfigure}
    \hfill
    \begin{subfigure}[b]{0.32\textwidth}
 \includegraphics[width=\linewidth]{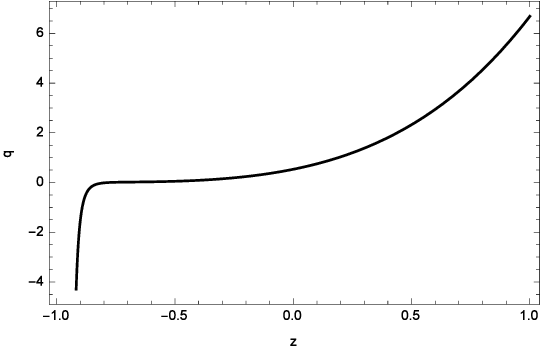}
        \caption{}
    \end{subfigure}
    \hfill
    \begin{subfigure}[b]{0.32\textwidth}
 \includegraphics[width=\linewidth]{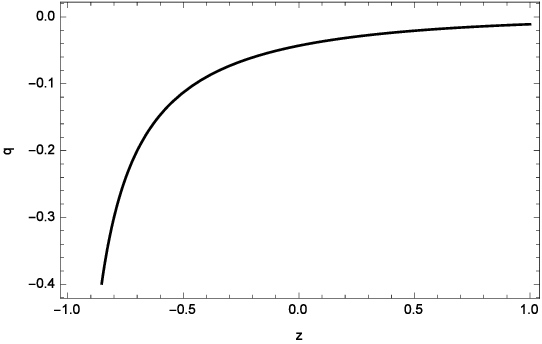}
        \caption{}
    \end{subfigure}
    
    \caption{deceleration parameter $q$ vs redshift $z$ for reconstructed $f(R)$ models  (a)Hubble Horizon, (b)Future Event Cutoff and (c)Granda-Oliveros }
    \label{q_vs_z}
    
\end{figure}

The deceleration parameter is a dimensionless quantity that predicts the expansion dynamics of the Universe by indicating whether the cosmic expansion is accelerating or decelerating. It is defined as
\begin{equation}
q=-1-\frac{\dot{H}}{H^{2}}.
\end{equation}
A positive value of the deceleration parameter $q>0$ corresponds to a decelerating Universe, whereas a negative value $q<0$ signifies an accelerating phase of cosmic expansion. The graph of the deceleration parameter $q(z)$ against redshift parameter $z$ is presented in  in Fig(\ref{q_vs_z}) for the reconstructed models with three different IR cutoffs.

In case of Hubble Horizon cutoff, the parameter values are $A_1 = -5, A_2 = 0.34,\delta = 1.222$ and $n = 3.56 $. From Fig:\ref{q_vs_z}(a), we observe that the trajectory of $q$  yields the transition from decelerating to accelerating phase, crossing the de-Sitter limit $q=-1$. With Parameter $A_1 = 45, A_2 = -0.25,\delta = 1.56$ and $n = 0.5688 $ the graph of deceleration prameter is drawn in figure Fig:\ref{q_vs_z}(b) against $z$ for future event horizon cutoff and  similar transition is obtained. For Granda-Oliveros cutoff, the parameter values used are $A_1 = 0.5, A_2 = 0.0012,\delta = 3.56$ and $n = 3.56 $, along with $\alpha = 10$ and $\beta = 5$. From Fig:\ref{q_vs_z}(c) it can be seen that $q(z)$ remains negative for GO cut-off, depicting accelerated expansion throughout the evolution of the Universe.

\subsection{Variation of $f(R)$ against $R$ for different IR cutoff }
\begin{figure}[htbp]
    \centering
    \begin{subfigure}[b]{0.32\textwidth}
 \includegraphics[width=\linewidth]{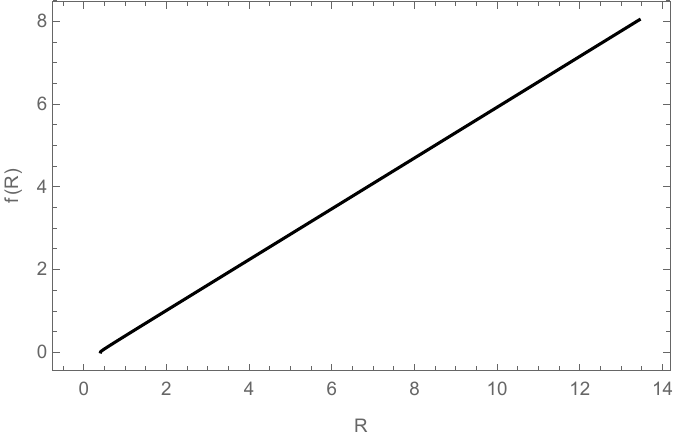}
        \caption{}
    \end{subfigure}
    \hfill
    \begin{subfigure}[b]{0.32\textwidth}
\includegraphics[width=\linewidth]{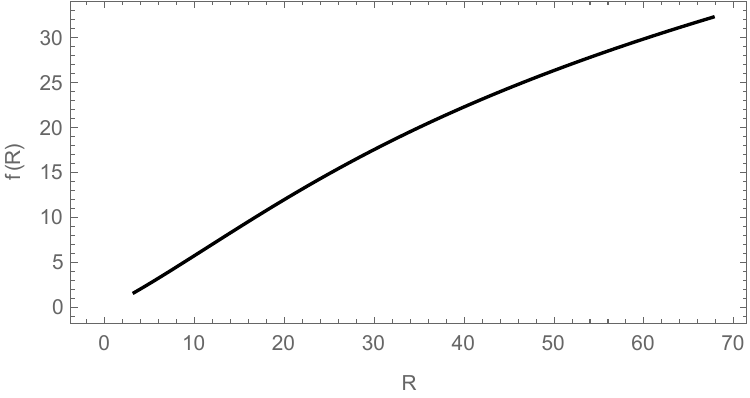}
        \caption{}
    \end{subfigure}
    \hfill
    \begin{subfigure}[b]{0.32\textwidth}
 \includegraphics[width=\linewidth]{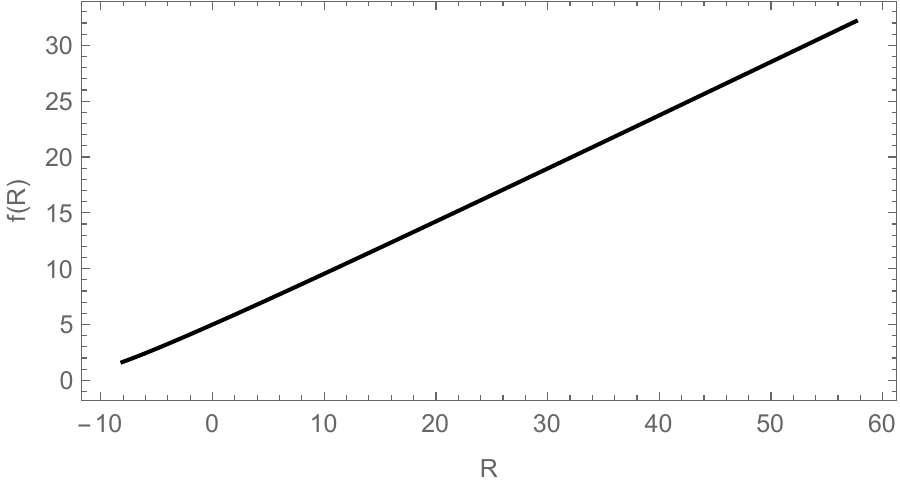}
        \caption{}
    \end{subfigure}
    
    \caption{$f(R)$ vs $R$ for reconstructed $f(R)$ models  (a)Hubble Horizon, (b)Future Event Cutoff and (c)Granda-Oliveros }
    \label{f_vs_R}
\end{figure}

The behavior of the reconstructed function $f(R)$ for the Hubble horizon, future event horizon, and Granda--Oliveros (GO) cutoffs is presented in Fig.~\ref{f_vs_R}. For the Hubble horizon case Fig:\ref{f_vs_R}(a), the function increases smoothly and almost linearly with the Ricci scalar $R$. In contrast, the future event horizon Fig:\ref{f_vs_R}(b) shows a slightly curved but steadily increasing profile. For the GO cutoff Fig:\ref{f_vs_R}(c), the function also exhibits a smooth monotonic increase, but with a comparatively steeper slope.
This smooth and positive progression
supports the physical plausibility and mathematical robustness for different reconstruction model.

\subsection{Viability Analysis $f(R)$ model under Different IR Cutoffs}
It is essential to examine the viability of any reconstructed model to ensure its consistency. Unlike Einstein’s General Relativity, where the gravitational action depends linearly on the Ricci scalar $R$, $f(R)$ gravity generalizes this dependence by replacing $R$ with an arbitrary function $f(R)$. Although such modifications provide a promising geometric explanation for the late-time accelerated expansion of the Universe without explicitly introducing dark energy, they must satisfy several physical conditions in order to remain reliable and realistic \cite{MAITY2025102184}. A viable $f(R)$ model should not only reproduce the observed cosmic acceleration but also remain free from instabilities at classical as well as quantum levels. Furthermore, it must be consistent with local gravity tests, particularly those performed within the solar system, where General Relativity has been verified to a high degree of precision. This requires that any deviation from standard gravity remains sufficiently small in high-density environments, while still allowing for significant effects at cosmological scales. These requirements impose strict conditions on the functional form of $f(R)$, which serve as important criteria for assessing the physical acceptability of the model \cite{Amendola2007}. The fundamental viability conditions for the reconstructed $f(R)$ model can be summarized as follows:\\

\textbf{(i) Positive Effective Gravitational Coupling:}  
The condition $f'(R) > 0$ must be satisfied to ensure that gravity remains attractive and to avoid ghost-like degrees of freedom in the theory. This requirement guarantees the positivity of the effective gravitational constant and prevents the emergence of negative-energy modes.\\

\textbf{(ii) Stability Condition:}  
For the model to remain stable under cosmological perturbations and to avoid tachyonic instabilities, it is necessary that $f''(R) > 0$. This condition also ensures the positivity of the scalar degree of freedom associated with modified gravity.\\

\textbf{(iii) Solar System Constraints:}  
In order to be consistent with local gravity tests, the model must satisfy the conditions
\[
f'(R) - 1 \ll 1, \quad R f''(R) \ll 1.
\]
These constraints ensure that deviations from General Relativity remain suppressed in high-density regions through mechanisms such as the chameleon effect, thereby preserving agreement with solar system observations \cite{Guo2014}.\\

For the Hubble horizon, future event horizon, and Granda--Oliveros (GO) cutoffs, the parameter values are chosen as $(A_1, A_2, \delta, n) = (0.23, 0.34, 1.222, 1.56)$, $(21, 0.002, 1.56, 3.5688)$, and $(0.0005, -12, 3.56, 3.56)$ respectively, with $\alpha = 10$ and $\beta = 5$ in the GO case.

\begin{figure}[htbp]
    \centering
    \begin{subfigure}[b]{0.32\textwidth}
 \includegraphics[width=\linewidth]{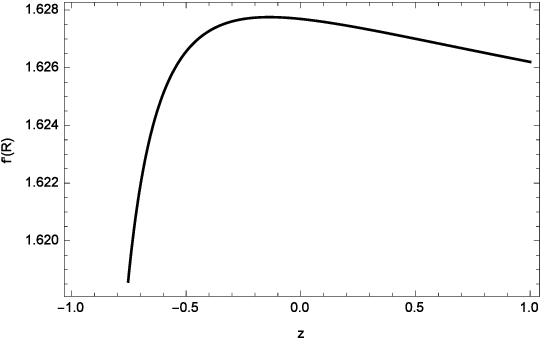}
        \caption{}
    \end{subfigure}
    \hfill
    \begin{subfigure}[b]{0.32\textwidth}
 \includegraphics[width=\linewidth]{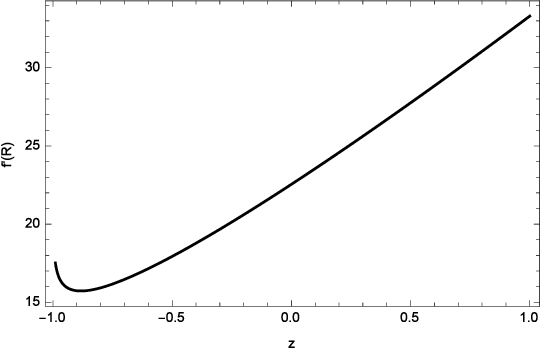}
        \caption{}
    \end{subfigure}
    \hfill
    \begin{subfigure}[b]{0.32\textwidth}
 \includegraphics[width=\linewidth]{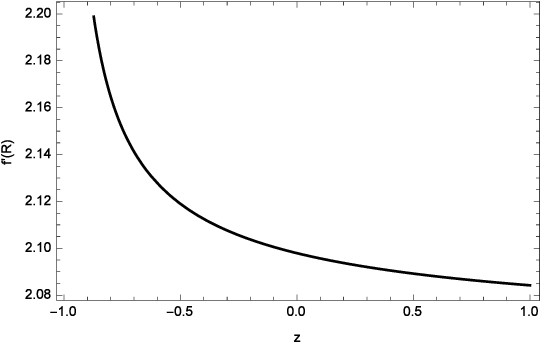}
        \caption{}
    \end{subfigure}
    
    \caption{$f'(R)$ vs $z$ for reconstructed $f(R)$ models  (a)Hubble Horizon, (b)Future Event Cutoff and (c)Granda-Oliveros }
    \label{f1_vs_z}    
\end{figure}

The behavior of the first derivative $f'(R)$ for all three cutoffs, as shown in Fig.~\ref{f1_vs_z}, remains positive throughout the considered redshift range. This confirms that the reconstructed models are free from ghost instabilities and that the effective gravitational coupling remains physically acceptable. Although the overall trend is similar for all cutoffs, the future event horizon case shows a relatively smoother and more stable evolution, while the GO cutoff exhibits a slightly steeper variation.\\

\begin{figure}[htbp]
    \centering
    \begin{subfigure}[b]{0.32\textwidth}
        \includegraphics[width=\linewidth]{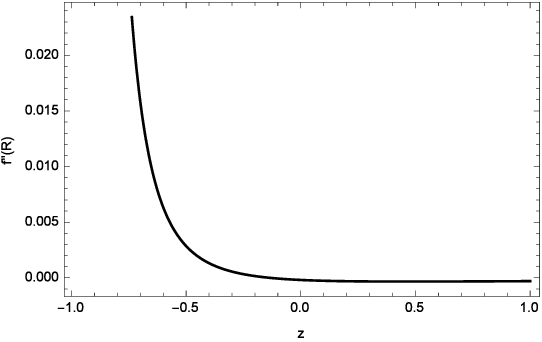}
        \caption{}
    \end{subfigure}
    \hfill
    \begin{subfigure}[b]{0.32\textwidth}
    \includegraphics[width=\linewidth]{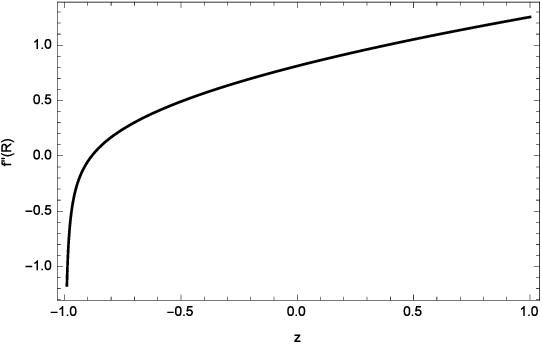}
        \caption{}
    \end{subfigure}
    \hfill
    \begin{subfigure}[b]{0.32\textwidth}
    \includegraphics[width=\linewidth]{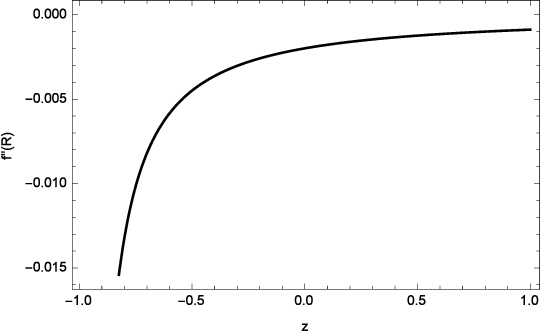}
        \caption{}
    \end{subfigure}
    
    \caption{$f''(R)$ vs $z$ for reconstructed $f(R)$ models  (a)Hubble Horizon, (b)Future Event Cutoff and (c)Granda-Oliveros }
    \label{f2_vs_z}    
\end{figure}

The second derivative $f''(R)$, shown in Fig.~\ref{f2_vs_z}, plays a crucial role in determining the stability of the model under cosmological perturbations. It is observed that for all three cutoffs, $f''(R)$ remains positive over the relevant range, indicating that the models satisfy the stability condition and do not suffer from tachyonic instabilities. The future event horizon model again shows a consistently positive and smooth behavior, whereas the Hubble and GO cases display mild variations but remain within the stable regime.\\
\begin{figure}[htbp]
    \centering
    \begin{subfigure}[b]{0.32\textwidth}
        \includegraphics[width=\linewidth]{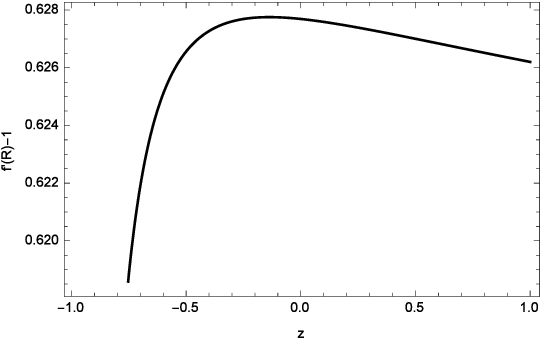}
        \caption{}
    \end{subfigure}
    \hfill
    \begin{subfigure}[b]{0.32\textwidth}
        \includegraphics[width=\linewidth]{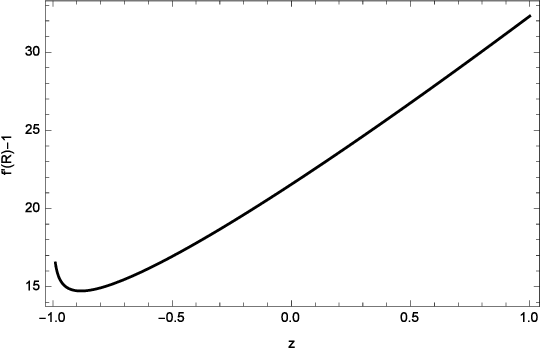}
        \caption{}
    \end{subfigure}
    \hfill
    \begin{subfigure}[b]{0.32\textwidth}
        \includegraphics[width=\linewidth]{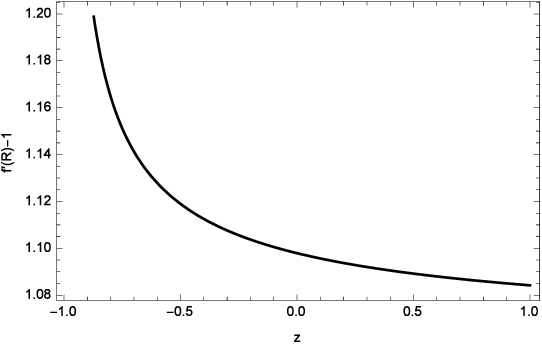}
        \caption{}
    \end{subfigure}
    
    \caption{$f'(R)-1$ vs $z$ for reconstructed $f(R)$ models  (a)Hubble Horizon, (b)Future Event Cutoff and (c)Granda-Oliveros }
    \label{f1-1_vs_z}    
\end{figure}

\begin{figure}[htbp]
    \centering
    \begin{subfigure}[b]{0.32\textwidth}
        \includegraphics[width=\linewidth]{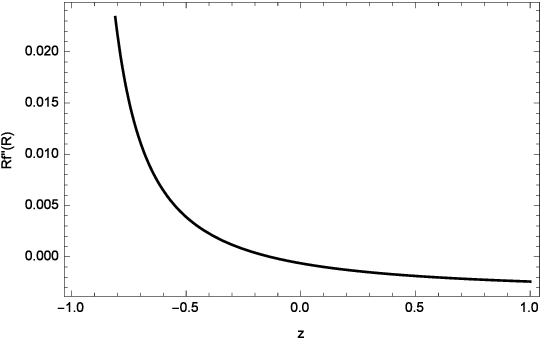}
        \caption{}
    \end{subfigure}
    \hfill
    \begin{subfigure}[b]{0.32\textwidth}
        \includegraphics[width=\linewidth]{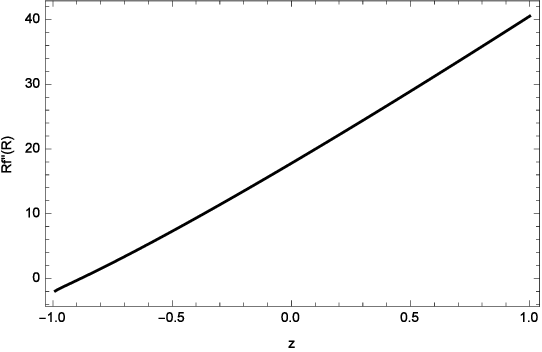}
        \caption{}
    \end{subfigure}
    \hfill
    \begin{subfigure}[b]{0.32\textwidth}
        \includegraphics[width=\linewidth]{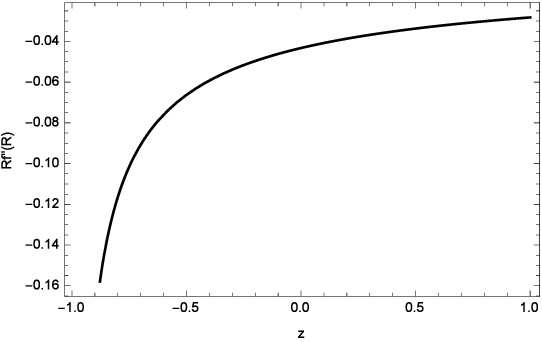}
        \caption{}
    \end{subfigure}
    
    \caption{$Rf''(R)$ vs $z$ for reconstructed $f(R)$ models  (a)Hubble Horizon, (b)Future Event Cutoff and (c)Granda-Oliveros }
    \label{Rf2_vs_z}    
\end{figure}

To further examine the local gravity constraints, the quantities $|f'(R) - 1|$ and $|R f''(R)|$ are analyzed, as shown in Fig.~\ref{f1-1_vs_z} and Fig.~\ref{Rf2_vs_z}. For all three cutoffs, these quantities are found to be significantly small across the considered redshift range. This indicates that deviations from General Relativity are well suppressed in high-density environments, ensuring consistency with solar system observations.\\

Overall, the viability analysis demonstrates that all three reconstructed $f(R)$ models satisfy the essential physical requirements. They successfully reproduce late-time cosmic acceleration while remaining free from ghost and tachyonic instabilities, and they are consistent with solar system constraints. Among the three, the future event horizon cutoff exhibits the most stable and smooth behavior, while the Hubble and GO cutoffs also remain viable within the chosen parameter ranges.
\section{Conclusion}

In this study, the reconstruction of the $f(R)$ gravity model has been carried out within the framework of Viaggiu holographic dark energy model for unification of the energy-based and geometry-based approaches to explain the observed late-time cosmic acceleration. Here we have considered three different infrared cutoffs, namely the Hubble horizon, future event horizon and Granda–Oliveros cutoffs for our investigation. The analytical forms of $f(R)$ are obtained for each case provide a clear understanding of how the choice of cutoff influences the behavior of the reconstructed model. The cosmological analysis of the obtained models are done via graphical representations of the cosmographic parameters namely equation of state parameter and deceleration parameter. It is found that for the Hubble Horizon cutoff  the EoS parameter shows a quintom-like behavior. In case of Future Event Horizon parameter values EoS parameter presents a transition from the dust dominant era to quintessence and enters the phantom region for the decreasing redshift parameter $z$. For Granda--Oliveros Cutoff, the graph of EoS parameter depicts only the quintessence kind of behaviour throughout the evolution.
From the graphs of deceleration parameter against the redshift parameter it is observed that in case of Hubble Horizon cutoff and future event horizon cutoff, the trajectory of $q$ yields transition from decelerating to accelerating phase and for Granda-Oliveros cutoff $q(z)$ remains negative indicating accelerated expansion throughout the entire evolution of the Universe. Variation of $f(R)$ with respect to $R$, shows that the reconstructed model is capable of describing the evolution of the Universe from a decelerated phase to the present accelerated phase. The graphical behavior further supports the physical viability and consistency of the model. The viability analysis further confirms the physical consistency of the reconstructed models. For all three infrared cutt offs, the conditions $(f'(R))>0$ and $(f''(R)>0)$ are satisfied throughout the considered redshift range, ensuring absence of ghost and tachyonic instabilities. Also, the quantities $(f'(R)-1)$ and $(Rf''(R))$ remain sufficiently small, indicating compatibility with local gravity and solar system constraints. Though all the three reconstructed models satisfy the required criteria, the future event horizon exhibits comparatively smoother and more stable over the evolution. Overall, the results indicate that the reconstructed $f(R)$ model provides a realistic and flexible framework for explaining cosmic acceleration, while remaining consistent with known cosmological behavior.

\bibliographystyle{elsarticle-num}
\bibliography{reference}
\end{document}